\documentclass[12pt]{iopart}
\usepackage{amssymb}
\usepackage{graphicx}
\usepackage{harvard}

\def\K{{\rm K}} 
\def\s{{\rm s}} 
\def\yr{{\rm yr}} 
\def\Gyr{{\rm G}\yr} 
\def\Hz{{\rm Hz}} 
\def\m{{\rm m}} 
\def\mum{\mu\m} 
\def\cm{{\rm c}\m} 
\def\pc{{\rm pc}} 
\def\kpc{{\rm k}\pc} 
\def\Ms{M_\odot} 
\def\erg{{\rm erg}} 

\def\d{{\rm d}}
\def\e{{\rm e}}
\def\SgrA{{Sgr A*}}

\begin{document}

\title[Observational limits on gravastars]{Where are all the gravastars?  Limits upon the gravastar model
  from accreting black holes}

\author{Avery E. Broderick$^1$ \& Ramesh Narayan$^1$}
\address{$^1$ Institute for Theory and Computation,
  Harvard-Smithsonian Center for Astrophysics, MS 51, 60 Garden
  Street, Cambridge, MA 02138, USA}
\ead{abroderick@cfa.harvard.edu, rnarayan@cfa.harvard.edu}

\begin{abstract}
The gravastar model, which postulates a strongly correlated thin shell
of anisotropic matter surrounding a region of anti-de Sitter space,
has been proposed as an alternative to black holes.  We discuss
constraints that present-day observations of well-known black hole
candidates place on this model.  We focus upon two black hole candidates known to
have extraordinarily low luminosities: the supermassive black hole
in the Galactic Center, Sagittarius A*, and the stellar-mass black
hole, XTE J1118+480.  We find that the length scale for modifications
of the type discussed in Chapline et al. (2003) must be sub-Planckian.
\end{abstract}


\maketitle

\section{Introduction}

There is strong observational evidence that many massive stars end
their lives as dark, compact objects with masses larger than $3\,\Ms$\cite{Nara:05}.
In the context of general relativity, the only possible stable
configuration for such an object is a black hole.  Recently,
\citeasnoun{Brod-Nara:06} demonstrated that the current near-infrared flux
limits on the black hole candidate Sagittarius A* (Sgr A*) at the
Galactic Center imply the existence of a horizon in this object.
Their argument assumed only that any non-horizon model has reached
steady state under continued accretion and that general relativity is
an acceptable description of gravity outside the alternative object's
photosphere.

In recent years a class of alternatives to general relativity has been
advanced in which a large-scale quantum phase transition occurs during
stellar collapse, preventing the formation of a horizon
\cite{Mazu-Mott:01,Chap_etal:03,Viss-Wilt:04,Cart:05,Lobo:06}.  In the most
developed of these ``gravastar'' models, general relativity is an
emergent theory failing at small length scales.
Necessarily, these models are characterized by a large surface
red-shift, enabling them to mimic black holes.

In principle, the gravastar model can escape the argument presented in
\citeasnoun{Brod-Nara:06} by possessing a large thermal capacity
\cite{Chap_etal:03} and therefore not reaching steady state.
However, it is expected that most compact objects in astrophysics grow
substantially in mass via accretion.  Therefore, if some of these are
gravastars, there is a natural mechanism for providing large amounts
of energy to heat them up.  Indeed, stellar mass black holes are
expected to accrete substantial fractions of their mass during
formation in the form of a fallback disk 
\citeaffixed{Woos-Hege-Weav:02,Frye-Hege-Lang-Well:02}{see, e.g.,}.  Similarly, within
the context of the standard heirarchical models of galaxy formation,
supermassive black holes are expected to grow primarily by accretion
\citeaffixed{Solt:82,Yu-Trem:02,DiMa-Spri-Hern:05}{see, e.g.,}.

Here we discuss the constraints that may be placed upon the gravastar
models by observations of black hole candidates.  For the purpose of
concreteness we restrict ourselves to the particular model presented by
\citeasnoun{Chap_etal:03}, but the arguments may be applied to dark energy
stars generally.  In sections \ref{TE} and \ref{OC} we discuss the
thermal evolution of gravastars and present the observational
constraints placed upon the parameters of the \citeasnoun{Chap_etal:03} gravastar model,
respectively.  Concluding remarks are contained in \ref{D}.

\section{Thermal Evolution} \label{TE}
\subsection{Accretion Heating}
The heating of neutron stars via accretion has been well documented.
In contrast, there is little evidence for similar heating in the case
of black-hole candidates.  Indeed, this fact has been used to argue
for the presence of event horizons in these objects
\cite{McCl-Nara-Rybi:04,Brod-Nara:06}.
However, the absence of detectable heating may also be consistent with
a gravastar if the heat capacity is large enough that it requires
prodigious amounts of heat to produce small changes in temperature.
Nevertheless, some level of accretion heating is unavoidable and, as
we show, provides surprisingly strong constraints on the gravastar
model.

In the model described in \citeasnoun{Chap_etal:03}, general
relativity is an emergent theory of an underlying Lorentz violating
microphysical theory, failing at small lengthscales.  In this context,
the gravastar is the result of a BEC-like phase transition induced by strong
gravity.  The internal energy $U$ of such a gravastar is related to its
mass $m = M/M_\odot$ in solar units and temperature $T$ by
\begin{equation}
U = \mathcal{U} \xi^{-1} m^3 T^3\,,
\label{gravastar_energy}
\end{equation}
where $\mathcal{U} \simeq 5.8\times10^{34}\,\erg\ K^{-3}$, and $\xi =
l/l_{\rm pl}$ is the length scale in Planck units at which general
relativity fails to adequately describe gravity. 
Note that while the particular form of $U$ is dependent upon the
microphysics underlying a gravastar or dark-energy star, given a
postulated form of $U$ the argument described below will necessarily
constrain the given model.

The large surface redshifts employed in gravastar models
have two immediate implications: (i) the internal energy generated per
unit rest mass accreted is very nearly $c^2$, i.e., the usual
accretion efficiency factor $\eta \approx 1$ (and will be neglected henceforth), and (ii) the radiation
emitted from the surface of the object should be almost a perfect
blackbody \cite{Brod-Nara:06}.  The energy evolution of an accreting
gravastar is determined by
\begin{equation}
\frac{dU}{dt} =  \dot{M} c^2 - L\,,
\end{equation}
where $\dot{M}$ is the mass accretion rate and $L$ is the total
luminosity.  Thus, from equation
(\ref{gravastar_energy}), we estimate that a gravastar which starts at
zero temperature and rapidly accretes a mass $\Delta m M_\odot$ will
be heated to a temperature as observed at infinity of
\begin{equation}
T_{\rm h}
\simeq
\left( \frac{\Ms c^2}{\mathcal{U} m^2} \right)^{1/3} \xi^{1/3}
\left( \frac{\Delta m}{m} \right)^{1/3}
\simeq
3.1\times10^6 m^{-2/3} \xi^{1/3}
\left( \frac{\Delta m}{m} \right)^{1/3}\,\K\,,
\label{heating}
\end{equation}
where $m$ is the final mass.  Here, ``rapid accretion'' means that the
rate at which internal energy is added via accretion is much larger
than the radiative luminosity $L$ of the heated surface, i.e., the
mass accretion rate $\dot{M}$ satisfies
\begin{equation}
\dot{M}
\gg
\frac{L}{ c^2}
=
\frac{g \mathcal{A} \sigma}{ c^2} m^2 T_{\rm h}^4 
\simeq
1.1\times10^{-3} m^{-5/3} \xi^{4/3}
\left(\frac{\Delta m}{m}\right)^{4/3} \dot{M}_{\rm Edd}\,,
\label{mdot_limit}
\end{equation}
where $\mathcal{A} m^2 = 108 \pi (G \Ms/c^2)^2$ is the effective area of
the radiating surface as measured at infinity,
$g=29/8$ is a degeneracy factor (3 types of neutrinos + photons,
assuming that the temperature in the local frame of the radiating
surface is high enough for neutrinos to be emitted in thermal
equilibrium),
and
$\dot{M}_{\rm Edd}=2.3\times10^{-9} m\, \Ms/\yr$
is the Eddington mass accretion rate, typically the maximum possible
in astrophysical systems.

In the opposite limit of slow accretion, the temperature is given by
the equilibrium or steady state value:
\begin{equation}
T_{\rm eq}
\simeq
\left[ \frac{ \dot{M} c^2}{ g \mathcal{A} \sigma m^2 } \right]^{1/4}
\simeq
1.7\times10^{7} \left(\frac{\dot{M}}{\dot{M}_{\rm Edd}}\right)^{1/4}
m^{-1/4} \,\K\,,
\label{slow_accretion}
\end{equation}
where $\sigma$ is the Stefan-Boltzmann constant.

\subsection{Radiative Cooling}
Unless we find a gravastar that is currently accreting in steady state, it is
generally insufficient to know the temperature to which accretion can
heat it.  Once accretion ceases, the surface will begin to cool via
thermal emission and the temperature will evolve according to
\begin{equation}
\frac{\d T}{\d t}
=
- \frac{L}{\d U/\d T}
=
- \frac{g \mathcal{A}\sigma}{3\mathcal{U}} \xi m^{-1} T^2\,.
\end{equation}
Therefore, the temperature is given by
\begin{equation}
T_{\rm c} = T_0 \left[ 1 + \frac{g \mathcal{A}\sigma T_0}{3\mathcal{U} \xi^{-1} m} t \right]^{-1}\,,
\label{cooling}
\end{equation}
where $T_0$ and $t$ are the temperature at which, and time since, radiative
cooling began.  This gives a typical cooling timescale for gravastars
of 
\begin{equation}
t_{\rm cool}
=
\frac{3\mathcal{U} \xi^{-1} m}{g \mathcal{A}\sigma T_0}
=
3.5\times10^{18} \frac{m}{\xi T_0} \,\yr\,,
\label{t_cool}
\end{equation}
which is generally quite long, unless $\xi$ is very large.

\section{Observational Constraints upon $\xi$}\label{OC}
The observational constraints we derive for $\xi$ arise from upper
limits on the temperature, $T$, of black hole candidates, obtained
through observations.  That is, if the gravastar model is a proper
description of the endpoint of gravitational collapse, it should
result in a thermally emitting surface whose emission is below all
limits set by spectral observations. Generally, each observed flux
$F_{\nu\,\rm obs}$ at frequency $\nu$ places a limit on
$T$ via the condition
\begin{equation}
F_{\nu\,\rm obs} >
F_\nu(T)
=
\frac{2 h \nu^3}{c^2}
\frac{\e^{-h\nu/kT}}{1-\e^{-h\nu/kT}} \frac{\mathcal{A}}{D^2}
m^2 \,,
\label{TFlimits}
\end{equation}
where $D$ is the distance to the compact mass.

As mentioned earlier, a putative gravastar would almost certainly have
acquired most of its mass via accretion, either as part of its birth
(e.g., during core collapse in a supernova explosion followed by the
rapid accretion of a fallback disk) or over an
extended period of time after birth (e.g., via accretion in a binary
system or as an active galactic nucleus).  Thus we expect $\Delta m/m
\sim 1$.

\begin{figure}[tb]
\begin{center}
\includegraphics[width=\columnwidth]{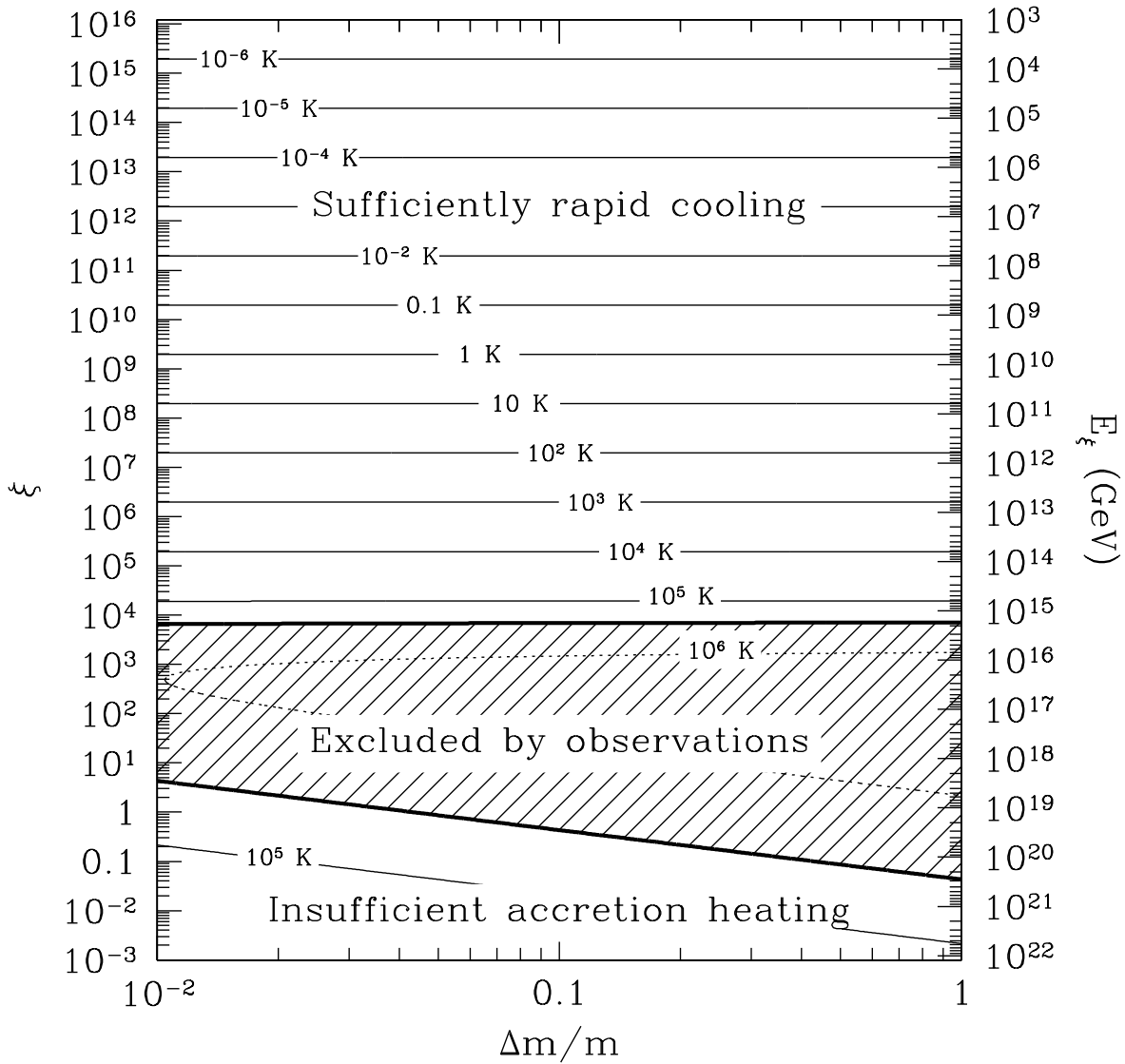} 
\end{center}
\caption{Contours of the predicted surface temperature today of a
gravastar of mass $8M_\odot$ heated by an episode of rapid accretion
$15\,$Gyr ago as a function of $\xi$ and the fractional mass accreted.
The hatched region of the plot is excluded by observations of the
black hole candidate XTE J1118+480.}
\label{fig:J1118}
\end{figure}

If most of the mass was accumulated via rapid accretion, the
temperature would have risen to the value given in equation
(\ref{heating}).  For any particular version of the gravastar model,
if this $T_{\rm h}$ is less than the observational limit $T_{\rm limit}$,
then the model is obviously consistent with observations.  Even if
$T_{\rm h}$ exceeds $T_{\rm limit}$, if the cooling time given in equation
(\ref{t_cool}) is shorter than the age of the system, which we may
take conservatively to be $t=15$ Gyr (the age of the universe), the
model is again consistent since the gravastar would have had time to
cool below $T_{\rm limit}$.  Therefore, as seen in Figure
\ref{fig:J1118}, the parameter space of the gravastar model is divided
into three regions corresponding to when (i) accretion heating is
insufficient to raise the temperature above detectable limits (labeled
as ``Insufficient accretion heating'' in Figures \ref{fig:J1118} \&
\ref{fig:SgrA}), (ii) accretion heating is sufficient to heat the
surface above, and  subsequent post-accretion cooling in $15\,$Gyr is
insufficient to reduce it below, detectable limits (labeled as
``Excluded by Observations''), and (iii) when post-accretion
cooling is sufficient to reduce the temperature below detectable
limits in $15\,\Gyr$ (labeled as ``Sufficiently rapid cooling'').
Region (ii) is clearly ruled out by the observational constraint.
We now apply this constraint via two black hole candidates with very
stringent flux limits.

\subsection{XTE J1118+480} \label{J1118}
\begin{table}
\caption{\label{table:J1118} UV and X-ray Flux Limits on XTE
  J1118+480.}
\begin{indented}
\item[]\begin{tabular}{cccc}
\br
$\nu\,(\Hz)$ & 
$F_\nu\,(\erg\cm^{-2}\s^{-1})$ &
$T_{\rm max}\,({\rm\K})$ &
Ref.\\
\mr
$2.3\times10^{15}$ & $2.4\times10^{-30}$ & $4.3\times10^5$ & \cite{McCl-Nara-Rybi:04}\\
$7.3\times10^{16}$ & $5.6\times10^{-32}$ & $2.7\times10^5$ & \cite{McCl-Nara-Rybi:04}\\
$2.5\times10^{17}$ & $1.4\times10^{-32}$ & $6.7\times10^5$ & \cite{McCl-Nara-Rybi:04}\\
\br
\end{tabular}
\end{indented}
\end{table}

The black hole candidate XTE J1118+480 is notable for its
extraordinarily low accretion luminosity \cite{McCl-Nara-Rybi:04},
which allows us to place sensitive limits upon any X-ray flux from the
vicinity of the horizon.  At a distance of $1.8\,\kpc$, the observed
flux limits place an upper limit upon the temperature (as measured at
infinity) of any horizon-sized surface of $T_{\rm limit} \leq
2.7\times10^{5}\,\K$ (see Table \ref{table:J1118}).  With a mass of
approximately $8\Ms$, this excludes the shaded region shown in Figure
\ref{fig:J1118}.  In particular, we can exclude $\xi$ of order unity
if as little as 4\% of the mass of the object was accreted some time
during its lifetime!  However, due to the low mass, and
correspondingly lower heat capacity (eq. \ref{gravastar_energy}), XTE
J1118+480 does not constrain values of $\xi$ above $\sim5\times10^{3}$.

\subsection{Sgr A*}
\begin{table}
\caption{\label{table:SgrA} Near-Infrared Flux Limits on \SgrA.}
\begin{indented}
\item[]\begin{tabular}{@{}llll}
\br\\
$\lambda\,(\mum)$ &
$F_\nu\,(\erg\cm^{-2}\s^{-1})$ &
$T_{\rm max}\,({\rm\K})$ &
Ref.\\
\mr\\
1.6 & $11$ & $2.4\times10^3$ & \cite{Stol_etal:03}\\
2.1 & $2.8$ & $1.6\times10^3$ & \cite{Ghez_etal:05b}\\
3.8 & $1.28$ & $1.1\times10^3$ & \cite{Ghez_etal:05b}\\
4.8 & $3.5$ & $1.7\times10^3$ & \cite{Clen_etal:04}\\
\br
\end{tabular}
\end{indented}
\end{table}

\begin{figure}[tb]
\begin{center}
\includegraphics[width=\columnwidth]{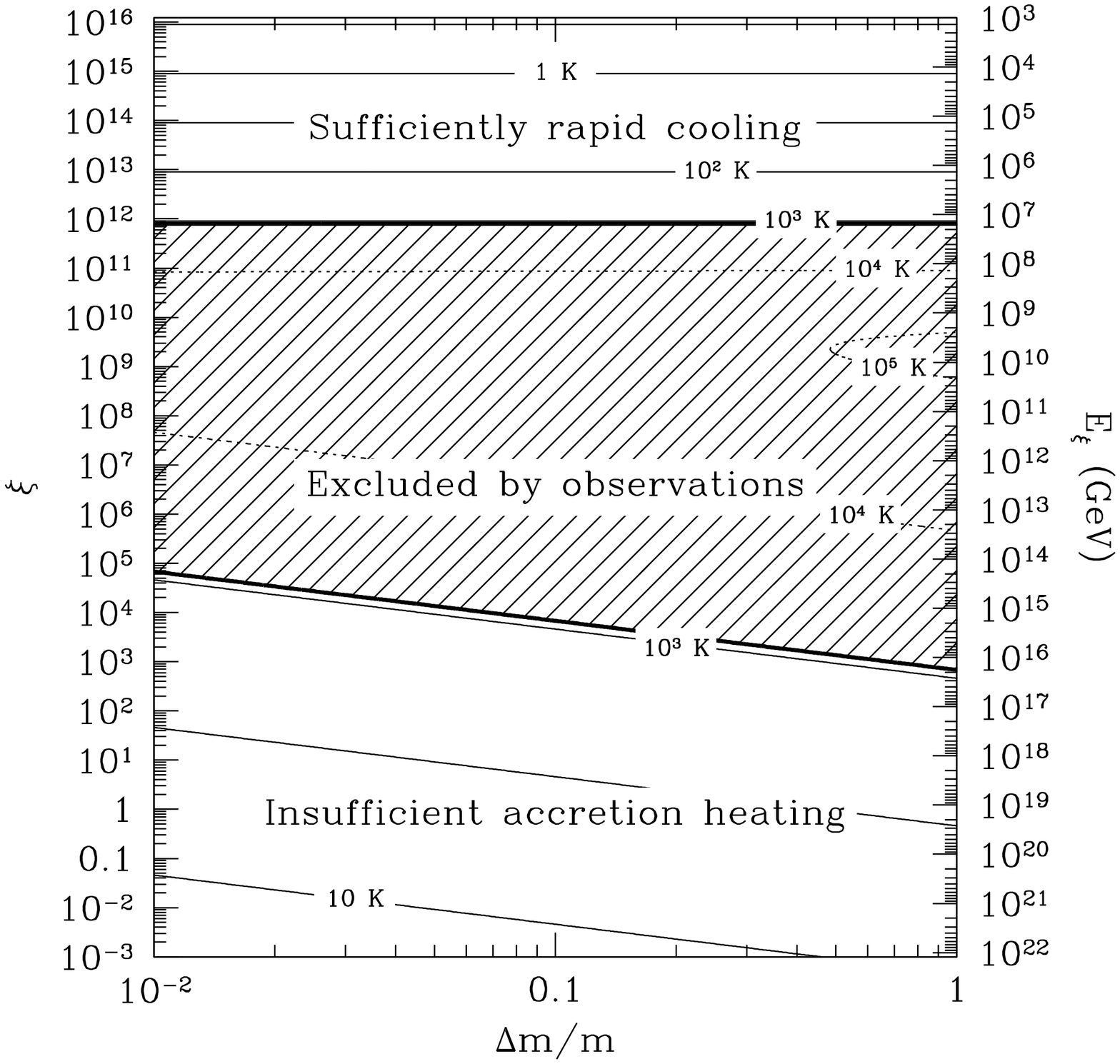} 
\end{center}
\caption{Contours of the predicted surface temperature today of a
gravastar of mass $3.7\times 10^6M_\odot$ heated by an episode of
rapid accretion $15\,$Gyr ago as a function of $\xi$ and the
fractional mass accreted.  The hatched region of the plot is
excluded by observations of the black hole candidate Sgr A*.}
\label{fig:SgrA}
\end{figure}

Being far more massive at $3.7\times10^6\Ms$
\cite{Scho_et_al:03,Ghez_et_al:05}, the accretion-heated temperature
of Sgr A* is necessarily lower than that for J1118+480.  Nevertheless,
the near-infrared (NIR) flux limits that have been placed upon Sgr A*
are sufficiently stringent to limit the brightness temperature of any
surface emission to less than $1.1\times10^{3}\,\K$ (Table
\ref{table:SgrA}).  As a consequence, values for $\xi$ between
approximately $10^4$ and $10^{11}$ are excluded (see Figure
\ref{fig:SgrA}) as long as Sgr A* has accreted a reasonable fraction
of its mass some time within the last $15\,\Gyr$.

\subsection{Sgr A*: Steady Accretion}
\begin{figure}[tb]
\begin{center}
\includegraphics[width=\columnwidth]{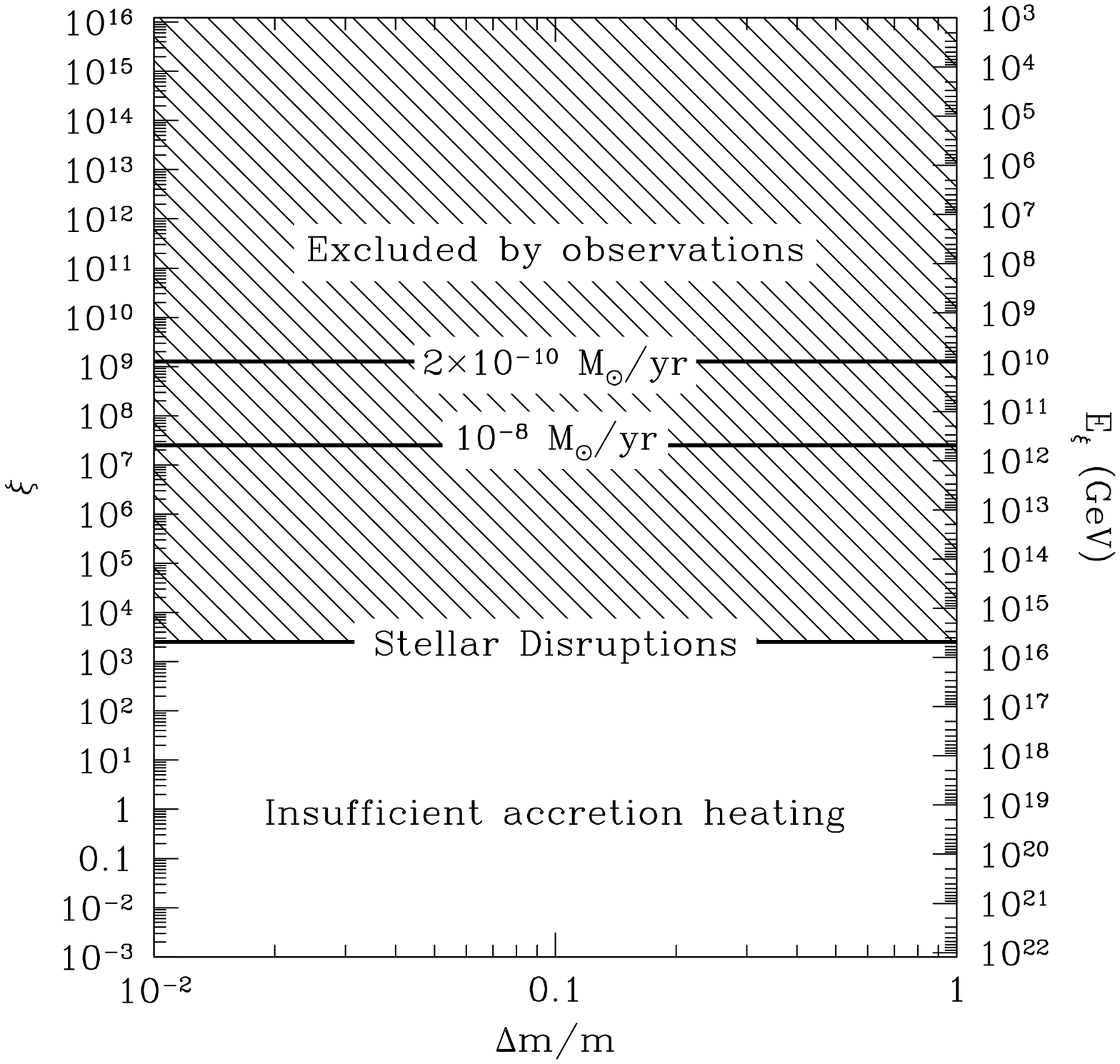}
\end{center}
\caption{Excluded values of $\xi$ as a result of the presently
  observed steady accretion onto Sgr A*, which for the purposes of
  this plot is assumed to have continued for the past $10\,\Gyr$.}
\label{fig:accretion}
\end{figure}

Additional constraints may be obtained from the fact that Sgr
A* is {\em presently}  accreting.  If the observed radio/sub-mm
luminosity of $10^{36} ~{\rm erg\,s^{-1}}$ of the source is accretion
powered with a canonical quasar radiative efficiency of 10\%, the
present accretion rate of gas must be at least
$2\times10^{-10}\,\Ms\yr^{-1}$ \citeaffixed{Brod-Nara:06}{see, e.g.,}.  In fact, typical radiatively
inefficient accretion flow models (RIAFs) imply accretion rates two
orders of magnitude higher
\citeaffixed{Nara-Yi-Maha:95,Yuan-Quat-Nara:03}{$\sim10^{-8}\,\Ms\yr^{-1}$, see, e.g.,}.
Finally, if stellar capture events are included the average accretion
rate can be as high as $10^{-5}$ to $10^{-3}\,\Ms\yr^{-1}$
\cite{Mago-Trem:99}.

Generally, the constraints that these accretion rates place upon $\xi$
depend on the time over which the rates have been maintained (which
translates to a given value of $\Delta m/m$).  A natural estimate for
this time scale (especially in the context of stellar captures) is the
age of the Galaxy, roughly $10\,\Gyr$, and thus implies $\Delta m/m$
of $5\times10^{-7}$, $3\times10^{-5}$ and $3\times10^{-1}$ for
$\dot{M}$ of $2\times10^{-10}\,\Ms\yr^{-1}$, $10^{-8}\,\Ms\yr^{-1}$
and $10^{-4}\,\Ms\yr^{-1}$, respectively.  Via equations
(\ref{heating}) \& (\ref{TFlimits}), this places lower limits upon
$\xi$ of approximately $10^{9}$, $3\times10^{7}$ and $2\times10^3$,
for minimal, RIAF and stellar capture accretion rates, respectively,
and are shown in Figure \ref{fig:accretion}.

For continuous accretion the gravastar will not have had an
opportunity to cool.  In contrast, stellar capture events produce
transient rapid accretion and allow the object to cool over the
intercapture timescales ($\sim10^4\,\yr$).  The upper limit implied by
efficient cooling in this case corresponds to $\xi\lesssim10^{18}$,
which is already ruled out by the continuous gas accretion even at the
minimal level.

\subsection{Total}
\begin{figure}[tb]
\begin{center}
\includegraphics[width=\columnwidth]{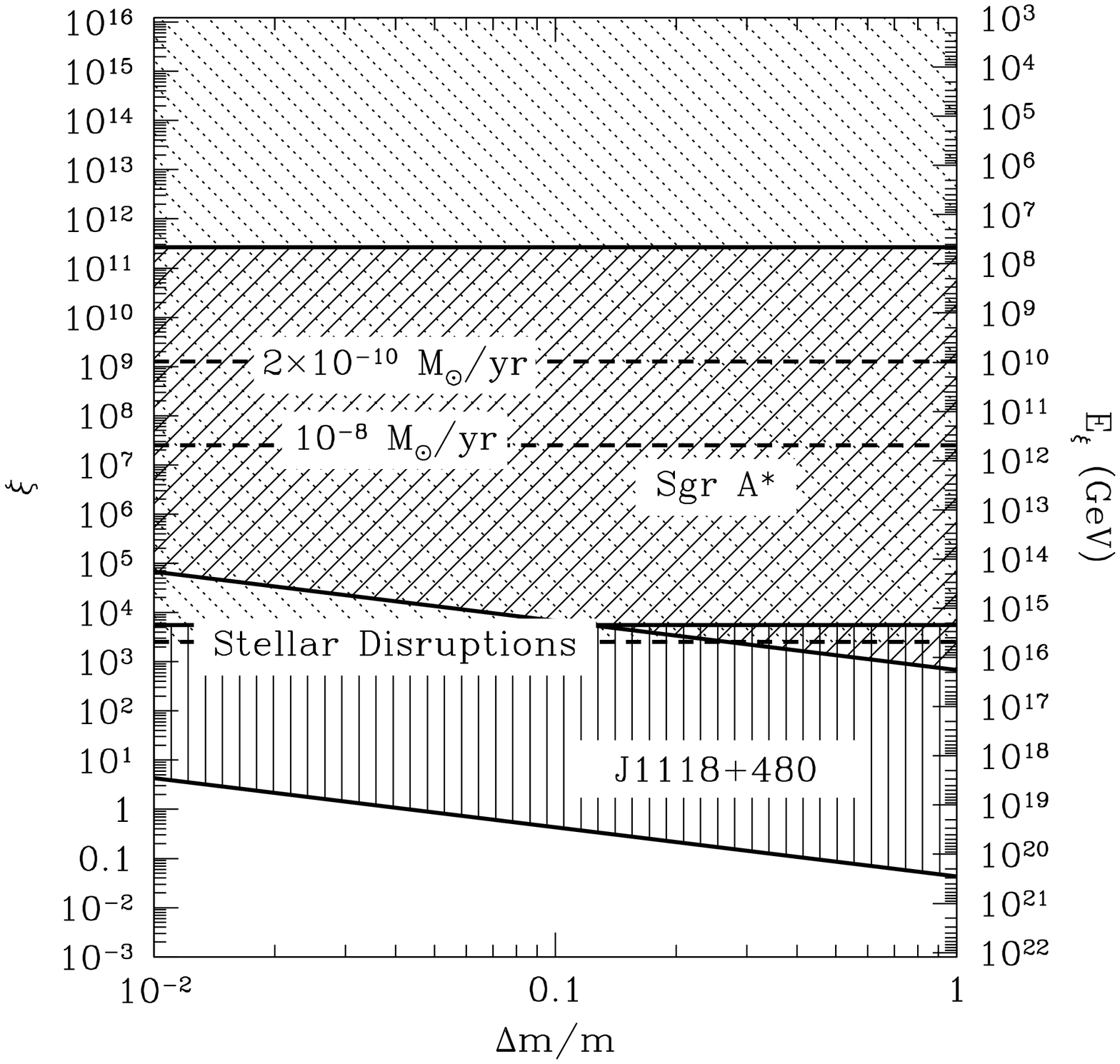} 
\end{center}
\caption{The currently excluded regions of the $\xi$--$\Delta m/m$
  parameter space. The vertically hatched region is excluded by observations
  of XTE J1118+480 (Figure \ref{fig:J1118}).  The solid-diagonal
  hatched region is excluded by observations of Sgr A* (Figure
  \ref{fig:SgrA}).  Finally, the the horizontal dashed lines show the lower
  limits placed upon $\xi$ by steady accretion onto Sgr A* (Figure
  \ref{fig:accretion}), which for the purposes of this plot is assumed
  to have continued for the past $10\,\Gyr$.}
\label{fig:total}
\end{figure}
The combined limits imposed by XTE J1118+480 and Sgr A* are shown in
Figure \ref{fig:total}.  We see that all values of $\xi$ larger than
unity are ruled out.  In other words, present-day astronomical
observations rule out modifications of general relativity of the kind
described by \citeasnoun{Chap_etal:03} on all scales larger than the Planck
length!

\section{Discussion} \label{D}
Due to the long-range correlations inherent in gravastar theories, the
heat-capacities of supermassive black holes will typically be
significantly larger than expected for normal hadronic matter.  As a
consequence, steady-state arguments of the kind presented in
\citeasnoun{Brod-Nara:06} may not be directly applicable.  Nevertheless, we
find that it is possible to strongly constrain the parameters of
gravastar models with essentially similar considerations.

A generic feature of gravastar theories is the necessity of a
spacetime phase-transition at large energies, required to facilitate
the creation of the exterior, strongly correlated matter shell and
interior anti-de Sitter space.  Given that the physics associated with
such a phase transition requires unknown modifications of general
relativity, it is difficult to rule out all possible variants of the
concept.  However, in the context of the most physically motivated
example to date, viz., the model of \citeasnoun{Chap_etal:03}, it is
possible to constrain quite strongly the parameters of the postulated
phase transition.  In particular, we find that the length scales over
which modifications are allowed must be smaller than the Planck length
at which, presumably, quantum gravitational effects enter in any case.

More generally, the lack of discernible thermal emission from the
surfaces of known black hole candidates provides a strong
observational constraint on any alternate gravitational theory that
does not admit horizons.

\ack
This work was supported in part by NASA grant
NNG04GL38G.  A.E.B. gratefully acknowledges the support of an ITC
Fellowship from Harvard College Observatory.

\newpage
\bibliography{glimits.bib}
\end{document}